\documentclass[useAMS,usenatbib]{mn2e}

\usepackage{graphicx}
\usepackage{natbib}

\usepackage{txfonts}
\title[M81 Group Extended Dust]{On the Origin of M81 Group Extended Dust Emission}
\author[J. I. Davies et al.]
{J. I. Davies$^{1}$, 
C. D. Wilson$^{15}$,
R. Auld$^{1}$, 
M. Baes$^{4}$, 
M. J. Barlow$^{3}$, 
G. J. Bendo$^{2}$, 
J. J. Bock$^{5}$, \newauthor
A. Boselli$^{3}$, 
M. Bradford$^{5}$, 
V. Buat$^{3}$, 
N. Castro-Rodriguez$^{6}$, 
P. Chanial$^{7}$, 
S. Charlot$^{8}$, 
L. Ciesla$^{3}$, \newauthor
D. L. Clements$^{2}$, 
A. Cooray$^{9}$, 
D. Cormier$^{7}$, 
L. Cortese$^{1}$, 
E. Dwek$^{10}$, 
S. A. Eales$^{1}$, \newauthor
D. Elbaz$^{7}$, 
M. Galametz$^{7}$,
F. Galliano$^{7}$, 
W. K. Gear$^{1}$, 
J. Glenn$^{11}$, 
H. L. Gomez$^{1}$, \newauthor
M. Griffin$^{1}$, 
S. Hony$^{7}$, 
K. G. Isaak$^{1, 12}$, 
L. R. Levenson$^{5}$, 
N. Lu$^{5}$, 
S. Madden$^{7}$, \newauthor
B. O'Halloran$^{2}$, 
K. Okumura$^{7}$, 
S. Oliver$^{13}$, 
M. J. Page$^{14}$, 
P. Panuzzo$^{7}$, 
A. Papageorgiou$^{4}$, \newauthor
T. J. Parkin$^{15}$, 
I. Perez-Fournon$^{6}$,
M. Pohlen$^{1}$, 
N. Rangwala$^{11}$, 
E. E. Rigby$^{16}$, 
H. Roussel$^{8}$,  \newauthor
A. Rykala$^{1}$, 
N. Sacchi$^{17}$, 
M. Sauvage$^{7}$, 
B. Schulz$^{18}$, 
M. R. P. Schirm$^{15}$, 
M. W. L. Smith$^{1}$, \newauthor
L. Spinoglio$^{17}$, 
J. A. Stevens$^{19}$, 
 S. Srinivasan$^{8}$,
M. Symeonidis$^{14}$, 
M. Trichas$^{2}$, 
M. Vaccari$^{20}$, \newauthor
L. Vigroux$^{8}$, 
H. Wozniak$^{21}$, 
G. S. Wright$^{22}$, 
W. W. Zeilinger$^{23}$. \\ 
$^{1}$School of Physics \& Astronomy, Cardiff University, Queens Buildings The Parade, Cardiff CF24 3AA, UK. \\
$^{2}$Astrophysics Group, Imperial College, Blackett Laboratory, Prince Consort Road, London SW7 2AZ, UK. \\
$^{3}$Laboratoire d'Astrophysique de Marseille, UMR6110 CNRS, 38 rue F.  
Joliot-Curie, F-13388 Marseille France. \\
$^{4}$Sterrenkundig Observatorium, Universiteit Gent, Krijgslaan 281 S9,  
B-9000 Gent, Belgium. \\	 
$^{5}$Jet Propulsion Laboratory, Pasadena, CA 91109, United States;  Dept. of Astronomy, California Institute of Technology, Pasadena,  
CA 91125, USA. \\
$^{6}$Instituto de Astrof\'isica de Canarias, v\'ia L\'actea S/N, E-38200 La  
Laguna, Spain. \\
$^{7}$CEA, Laboratoire AIM, Irfu/SAp, Orme des Merisiers, F-91191
Gif-sur-Yvette, France. \\
$^{8}$Institut d'Astrophysique de Paris, UMR7095 CNRS, Universit\'e Pierre  
\& Marie Curie, 98 bis Boulevard Arago, F-75014 Paris, France. \\
$^{9}$Dept. of Physics \& Astronomy, University of California, Irvine,
CA 92697, USA. \\	
$^{10}$Observational  Cosmology Lab, Code 665, NASA Goddard Space Flight Center Greenbelt, MD 20771, USA. \\
$^{11}$Dept. of Astrophysical \& Planetary Sciences, CASA CB-389,  
University of Colorado, Boulder, CO 80309, USA. \\
$^{12}$ ESA Astrophysics Missions Division, ESTEC, PO Box 299, 2200 AG Noordwijk, The Netherlands. \\		
$^{13}$Astronomy Centre, Dept. of Physics and Astronomy, University of Sussex, UK. \\
$^{14}$Mullard Space Science Laboratory, University College London, Holmbury St Mary, Dorking, Surrey RH5 6NT, UK. \\
$^{15}$Dept. of Physics \& Astronomy, McMaster University, Hamilton, Ontario, L8S 4M1, Canada. \\	
$^{16}$School of Physics \& Astronomy, University of Nottingham, University  Park, Nottingham NG7 2RD, UK. \\
$^{17}$Istituto di Fisica dello Spazio Interplanetario, INAF, Via del Fosso del Cavaliere 100, I-00133 Roma, Italy. \\
$^{18}$Infrared Processing \& Analysis Center, California Institute of Technology, Mail Code 100-22, 770 South Wilson Av, Pasadena, CA 91125, USA. \\		
$^{19}$Centre for Astrophysics Research, Science \& Technology Research Centre, University of Hertfordshire, College Lane, Herts AL10 9AB, UK. \\
$^{20}$University of Padova, Dept. of Astronomy, Vicolo Osservatorio  3, I-35122 Padova, Italy. \\		
$^{21}$Observatoire Astronomique de Strasbourg, UMR 7550 Universit\'e de Strasbourg - CNRS, 11, rue de l'Universit\'e, F-67000 Strasbourg, France. \\
$^{22}$Astronomy Technology Center, Royal Observatory Edinburgh, Edinburgh, EH9 3HJ, UK. \\
$^{23}$Institut f\"ur Astronomie, Universit\"at Wien, T\"urkenschanzstr. 17, A-1180 Wien, Austria. }

\begin{document}

\date{Original July 2010}

\pagerange{\pageref{firstpage}--\pageref{lastpage}} \pubyear{2010}

\maketitle

\label{firstpage}

\begin{abstract}
Galactic cirrus emission at far infrared wavelengths affects many extra-galactic observations. Separating this emission from that associated with extra-galactic objects is both important and difficult.  In this paper we discuss a particular case, the M81 group, and the identification of diffuse structures prominent in the infrared, but also detected at optical wavelengths. The origin of these structures has previously been controversial, ranging from them being the result of a past interaction between M81 and M82 or due to more local Galactic emission. We show that over of order a few arc minute scales the far-infrared (Herschel 250$\mu$m) emission correlates spatially very well with a particular narrow velocity (2-3 km s$^{-1}$) component of the Galactic HI. We find no evidence that any of the far-infrared emission associated with these features actually originates in the M81 group. Thus we infer that the associated diffuse optical emission must be due to galactic light back scattered off dust in our galaxy. Ultra-violet observations pick out young stellar associations around M81, but no detectable far-infrared emission. We consider in detail one of the Galactic cirrus features, finding that the far-infrared HI relation breaks down below arc minute scales and that at smaller scales there can be quite large dust temperature variations.
\end{abstract}


\section{Introduction} 
It is clear that galaxy-galaxy interactions play a major role in the evolution of primordial gas clouds into the spectacular galaxies we see today. Particularly important and striking interactions occur when the speed of interaction is well matched to the internal velocities of the stars and gas. Thus small galaxy groups can potentially provide just the environment for dramatic gravitational disturbance. Beyond the Local Group the closest example of this is in the environment around M81. The M81 group lies at a distance of about 3.6 Mpc (Karachentsev et al. 2004) with radial velocities ranging from -34 km s$^{-1}$ for M81 to 203 km s$^{-1}$ for M82 (Chynoweth et al. 2008), thus tidal disruption is highly likely. 

Living up to expectations M81 is known to be surrounded by debris which has previously been best delineated by its emission at 21cm (Yun et al. 1993), although diffuse emission at wavelengths from the optical to the far-infrared can also be readily detected. Previously others have explicitly described the diffuse optical emission from between M81 and M82 as arising from stars either deposited there during a M81/M82 collision or subsequently formed in the gas that was stripped during this event (Sun et al. 2005). Additional support for this comes from a number of recent studies that have identified individual stars way beyond the discs of the individual galaxies involved (Davidge, 2008; Williams et al. 2009; De Mello et al. 2008; Mouhcine and Ibata, 2010). Contrary to this a study by Sollima et al. (2010) concludes that these stars can only represent a small fraction of the optical emission with the majority of the light being due to back scattering from foreground dust in our Galaxy, dust associated with 21cm cirrus emission. By looking at various far-infrared luminosity ratios Sollima et al. also conclude that the majority of the far-infrared emission is not associated with M81 but is again from Galactic cirrus.

Deciding on this issue is not a new problem. The area of sky around M81 has long been known for diffuse emission at optical wavelengths. Sandage (1976) describes it as 'high-latitude reflection nebulosity illuminated by the Galactic plane'. Arp (1965) describes faint diffuse emission that he assigns to the M81 group and one of the most prominent features has become known as 'Arp's loop'(see fig 1). So the origin of this emission has been somewhat controversial for many years. Our attention was drawn to this problem because recent wide field Spitzer and Herschel data of other regions of the sky show extended filamentary structures that are best explained by thermal emission from dust above the plane of our Galaxy, i.e. the Galactic cirrus whose large scale structure was first defined at lower resolution by IRAS. The emission looping around and apparently connected to M81 appears very similar in shape to this cirrus emission, yet one might also expect structures just like this within a galaxy group. To shed new light on this problem we compare Herschel high resolution far-infrared observations with high spatial and velocity resolution 21cm data.

\begin{figure}
\centering
\includegraphics[scale=0.555]{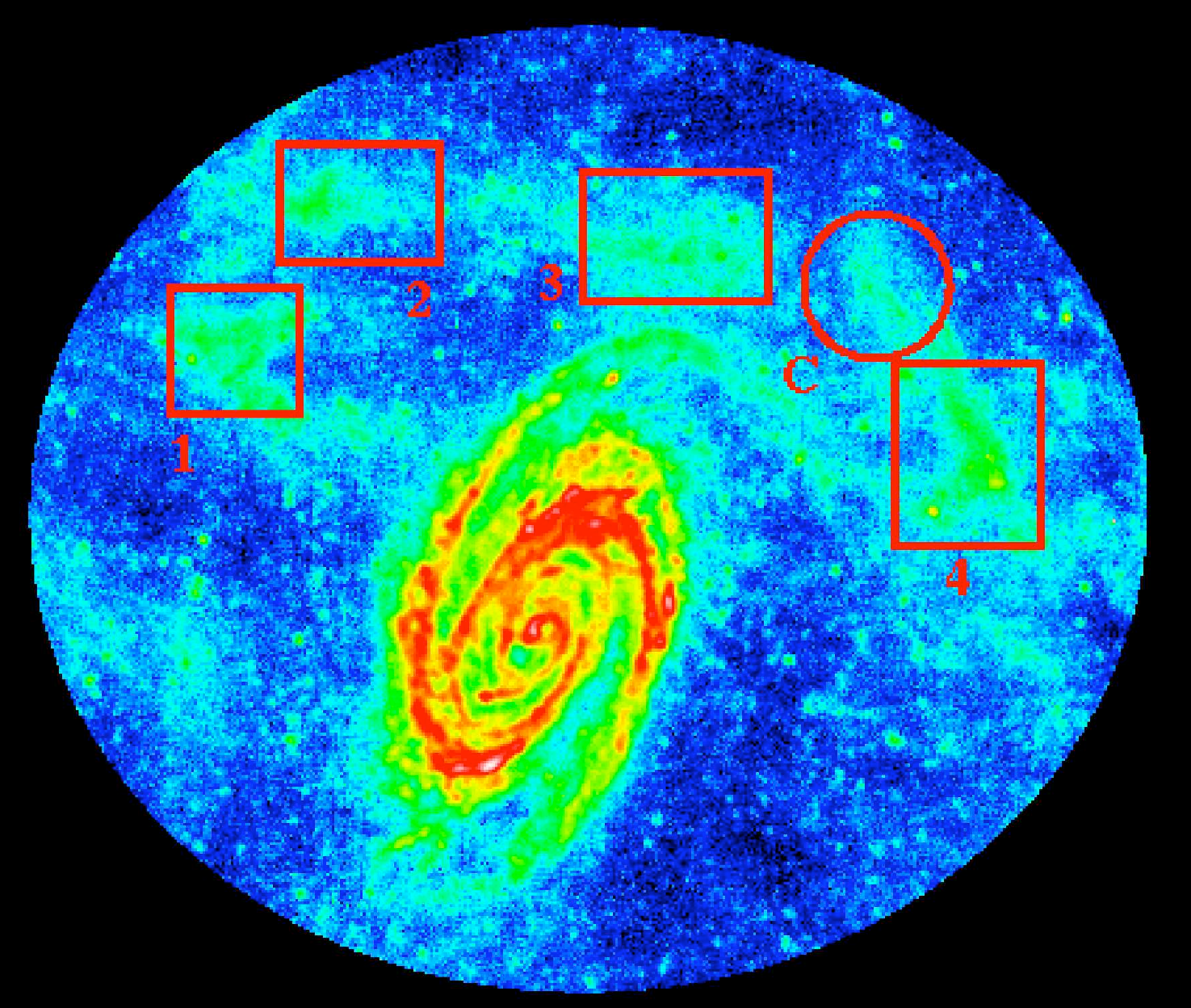}
\includegraphics[scale=0.58]{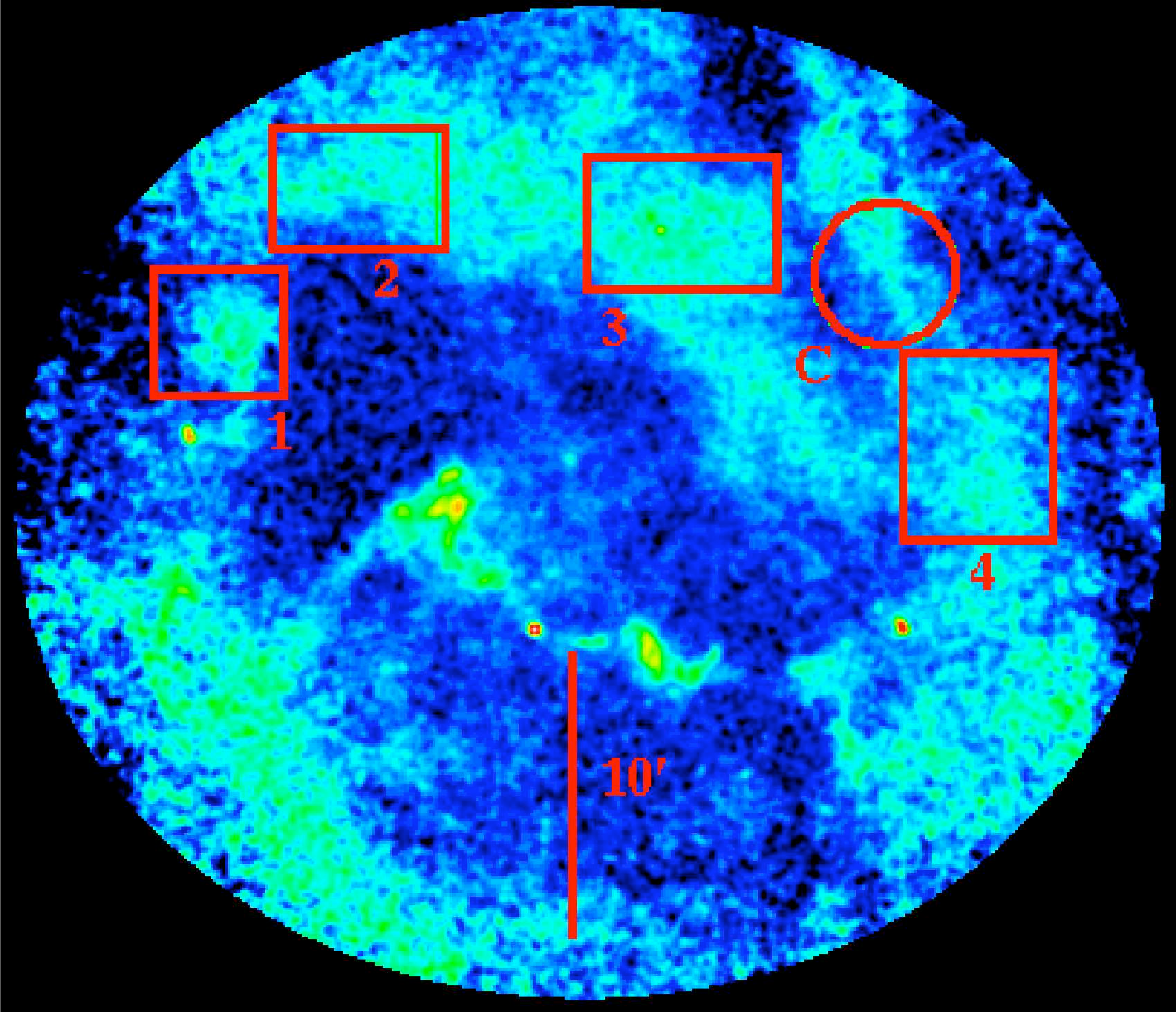}
\caption{
Top - the SPIRE 250$\mu$m data. Of particular interest is the extended emission to the north of M81, four particularly bright regions are highlighted by boxes 1 - 4. Arp's loop extends from M81 to box 1, box 2 and then box 3. A possible star formation region is highlighted by the circle and labelled C. Bottom - this is channel 75 from the THINGS M81 HI data cube which is centred at -1.2 km s$^{-1}$. The HI emission corresponds very well with the extended 250$\mu$m emission, again highlighted by the same boxes. A size of 10'$\approx10.5$ kpc at the distance of M81 is marked by the vertical line. North is towards the top and East to the left.}
\end{figure}

\section{Debris or cirrus?}
We will use two primary data sources to help us decide on the debris or cirrus issue: Herschel Space Telescope PACS 160$\mu$m (Poglitsch et al. 2010) and SPIRE (Griffin et al. 2010) 250, 350 and 500$\mu$m science demonstration scan map data and THINGS VLA data (Walter et al. 2008). 

PACS observations were performed as four pairs of orthogonal scans covering  an area of $\sim$40$\times$40 square arc min with a scan rate of 20''/sec.
The SPIRE observations consist of two repetitions of cross-linked scans over approximaetly the same area with a nominal scan speed of 30''/sec. The reduction of both the PACS and SPIRE data is described in Bendo et al. (2010) and references therein. The final rms of the images are $\sim$3, 12, 12 and 13 mJy/beam at 160, 250, 350 and 500$\mu$m. The full width at half maximum of the beams are $\sim$11, 18, 25 and 37 arcsec at 160, 250, 350 and 500$\mu$m respectively. There is a calibration uncertainty of $\approx15\%$ (Swinyard et al. 2010).
M81 was observed by the THINGS project using the Very Large Array (Walter et al. 2008). They have produced a 21cm data cube with a noise of $\sim$0.7mJy/beam, a spatial resolution of 12.5" and a velocity channel width of 2.6 km s$^{-1}$. 

Extended filamentary structures external to the disc of M81 are clearly seen in emission in all of the Herschel bands. In Fig 1 (top) we illustrate this with the 250$\mu$m data, four particularly bright regions are indicated by the red boxes. These same structures can also be clearly seen in the Spitzer (MIPS) 24, 70 and 160$\mu$m data (Sollima et al. 2010), so there is emission over a wide range of far infrared wavelengths. 

If this material is associated with M81 group tidal debris then it must consist of relatively large quantities of dust, and there must be a source of heating either from stars within the dust, also accounting for the optical emission, or from stars further away in M81. Alternately if it is Galactic cirrus then dust can account for all of the ultra-violet, optical and infra-red emission. In the optical it would be scattered light from the disc of our Galaxy and in the infra-red thermal emission.
We also have a deep optical SDSS image that clearly shows diffuse emission that is spatially correlated with the far-infrared (see section 4). Given that there is normally a correlation between Galactic cirrus gas and dust (Boulanger et al. 1996) the 21cm data may be useful in deciding this issue because in principle we can select the cirrus gas via its velocity.  We have smoothed the THINGS 21cm data to the spatial resolution of the SPIRE 250$\mu$m data, re-gridded to the same pixel scale and then considered the spatial correlation of the 21cm and 250$\mu$m data. The published THINGS moment zero map, which is the sum of 21cm emission over $\sim$250 km s$^{-1}$ does not correlate particularly well with the diffuse far infrared emission (see below), but that is not true of the individual channel maps.  The problem with M81 is that because of its relatively low velocity, its HI is mixed with that of the Galaxy and separating the two is not without ambiguity. M81 can be traced throughout the THINGS 21cm channel maps, but at velocities around zero a number of other striking and extended structures appear with very narrow velocity widths. For example Fig 1. (bottom) shows the HI single channel map centred at a velocity of -1.2 km s$^{-1}$, the structures seen in this image are almost completely contained within this one channel of velocity width 2.6 km s$^{-1}$ and disappear completely two channels away. What is striking is the correspondence of the HI emission in this narrow velocity range with the 250$\mu$m emission. If at the distance of M81, the HI structures would have to sustain an almost constant line of sight velocity over a distance of more than 30 kpc. This does not seem possible for what in this case would almost certainly be a tidal feature, which more typically have a velocity range of hundreds of km s$^{-1}$ (Minchin et al.2007). On the other hand, if it is cirrus above the plane of our Galaxy (distance at most 1 kpc?) then the feature has a length of only about 9 pc and its almost constant line of sight velocity is more understandable. In fact small velocity widths seem to be a characteristic of low velocity hydrogen clouds above the plane of our galaxy, these typically have velocity widths of 3-7 km s$^{-1}$ (Stanimirovic et al. 2006). 

Boulanger et al. (1996) have previously considered the correlation of Galactic HI with far-infrared emission over a wide range of wavelengths (see also Davies et al. 1997). They use COBE (DIRBE) far-infrared and Leiden/Dwingeloo HI data with a spatial resolution of about 40 arc min to derive a linear relation between the HI column density and the far-infrared emission (Fig 2). This spatial scale is large compared to the scale of our image, but nevertheless we have considered the relationship between the four regions labeled on Fig. 1 (each only a few arc minutes in size), these are marked as crosses on Fig. 2. The scatter in the Boulanger et al. data is about 1 MJy sr$^{-1}$ (see their Fig. 1) and we might expect even more scatter when averaging over our smaller regions. With the exception of box 1, to which we will return later, we conclude that the Herschel 250$\mu$m and THINGS HI data are consistent with the Boulanger et al. (1996) relation for Galactic cirrus emission.

\begin{figure}
\centering
\includegraphics[scale=0.5]{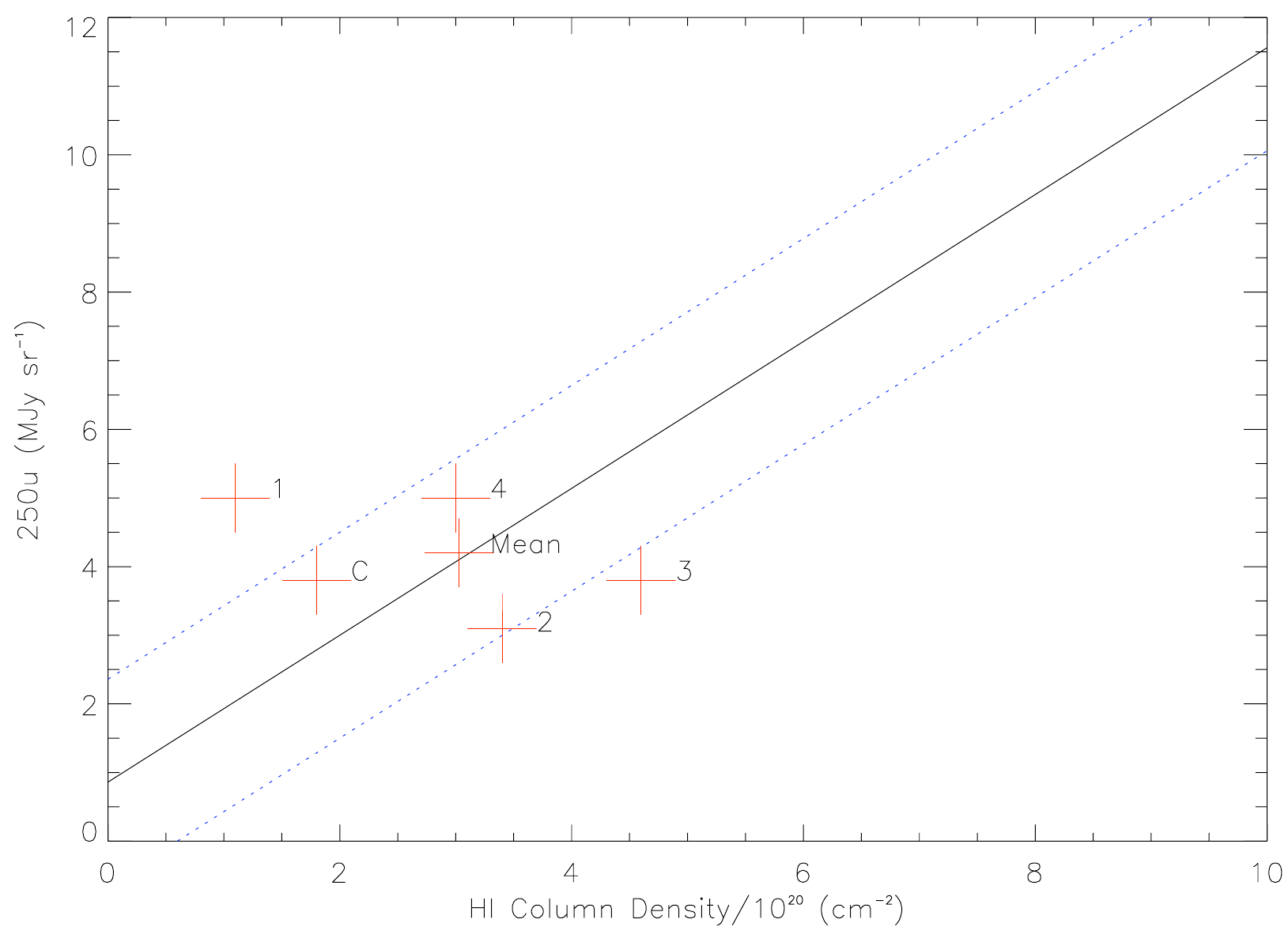}
\caption{The solid line gives the best fit to the Boulanger et al. (1996) DIRBE 240$\mu$m and Leiden/Dwingeloo HI data of FIR$_{240}$(MJy/sr$^{-1}$)=1.07N$_{HI}$/10$^{20}$(cm$^{-2}$)+0.86, the dotted lines indicate the scatter in the data. The four numbered points are the values for the corresponding rectangular apertures (HI in a single channel) shown on Fig. 1., their mean value is also marked by a red cross. The point marked C is for the circular aperture marked on Fig. 1 which covers the star formation region labeled C on Fig. 3.}
\end{figure}

After smoothing to the 500$\mu$m resolution and re-gridding we can use the 160, 250, 350 and 500$\mu$m data to estimate a temperature for the dust using a single temperature grey body with an emissivity that scales as $\lambda^{-2}$ (Li and Draine 2001). The derived temperatures for regions 1, 2, 3 and 4 are 20, 21, 18  and 21K respectively (each with an error of about $\pm$3K). Unhelpfully these temperatures are consistent with the temperature of Galactic cirrus, Boulanger et al. (1996) give a mean value of T$\approx$18K (see also 
del Burgo et al. 2003), and with the global dust temperature of M81 (Bendo et al. 2010).

Given the spatial similarities of the far-infrared emission and the Galactic atomic gas, the correspondence of the data with the Galactic far-infrared/HI relation and the similar temperature to Galactic cirrus we conclude that some substantial fraction of the SPIRE 250$\mu$m emission has an origin in our galaxy and not in the M81 group. 

\section{Is there any extended emission from the M81 group?}
Although it appears that some large fraction of the diffuse far-infrared emission is closely associated with Galactic cirrus there are enough differences between the far-infrared and the HI (discussed further in section 4) to make us curious about some of the emission features. There are for example other objects in this field that we might have hoped to detect in the far-infrared. There are three dwarf galaxies HoIX, A0952+69 and BK 3N and a large northern arm that is prominent in the HI and ultra-violet, but not in the far-infrared. Given that far-infrared and UV emission are both thought to be good tracers of star formation and that ultra-violet emission correlates well with high HI column density (Hibbard et al. 2005) ultra-violet and M81 HI data could provide further insight.

In Fig. 3 we show the GALEX FUV data and the THINGS HI data for the same region shown in Fig. 1. The HI data is now summed over the velocity range 15-180 km s$^{-1}$ to pick out hydrogen in the northern part of the M81 group. Note, this velocity range excludes the Galactic component shown in Fig. 1. The above three dwarf galaxies and the spiral arm are indicated on Fig. 3. All three dwarf galaxies are detected as strong sources in the ultra-violet and HI. Although they all appear to have young stars associated with them, none are unambiguously detected at 250$\mu$m or any of the other far-infrared wavelengths. In fact they all reside in regions where there is a distinct lack of far-infrared emission. Holmberg IX has previously been described as the 'Nearest Young Galaxy' (Sabbi et al. 2008). The nature of the stellar populations in BK3N and A0952+69 is less clear, but they are probably similar to that of Holmberg IX (Makarova et al. 2002). We can find no HI measurements for BK3N and A0952+69, but Holmberg IX has a measured HI mass of $3.3\times10^{8}$ M$_{\odot}$ (Swaters and Balcells, 2002). For a typical gas to dust ratio of 150 this would place it well above our detection limit of about $10^{4}$ M$_{\odot}$ of dust, making it both a low metalicity and low dust content galaxy. It has been suggested that all three dwarfs are tidal, having formed from the interaction of M81 and M82 some 10$^{8}$ years ago (Yun and Ho, 1994, Makarova et al. 2002). The dwarfs lack of far-infrared emission and low metalicity (De Mello et al. 2008) is surprising if they formed from gas originally in the disc of either M81 or M82. 

The region labeled C on Fig. 3 is prominent in both the ultra-violet and the M81 atomic hydrogen, it sits at the end of the northern spiral arm. This northern arm is a bit of an enigma because it is not seen in ground based optical data, though individual stars have been found using HST data (Williams et al. 2009). It seems that this is gas that has recently been drawn out of M81 and is only just starting to form stars. Coincident with the ultra-violet emission and morphologically very similar to it there is emission in the far-infrared (marked by a circle on Figs. 1 and 3). The problem associating this far-infrared emission with M81 is that there also appears to be Galactic cirrus emission in this region - it corresponds to a Galactic HI feature (Fig. 1, bottom). If there was far-infrared emission from this obviously active M81 star formation region then one might expect warmer dust, but its temperature, as derived above for the four other apertures, is 20$\pm$3K, consistent with Galactic cirrus. If we look at the relationship between the 250$\mu$m emission and the Galactic HI in this area (aperture C on Fig. 1) then it is perfectly consistent with the Galactic cirrus relation (Fig.2). Similar to the dwarf galaxies discussed above, there is no evidence that this star formation region prominent in both the ultra-violet and in atomic hydrogen at the velocity of M81 has any far-infrared emission associated with it.

The HI data (Fig 1) indicates that a sizeable fraction of the emission from region 1 is actually Galactic cirrus, but this region in particular does not fit the Boulanger et al. (1996) relation very well (Fig. 2) - there is excess far-infrared emission. Recently de Mello et al. (2008) have used HST ACS data to search for stars within Arp's Loop. Their data cover about a third of the northern part of region 1. Surprisingly they find both old ($>1$ Gyr) and young ($<10$ Myr) stars in the Loop. They suggest that the old stars were actually formed in the disc of M81 or M82 and were drawn out during a tidal passage, while the young stars have formed in situ. It is possible that these stars may be a heating source for dust that was also drawn out and so provide an explanation of the excess far-infrared emission from this region compared to expectations for the cirrus.  The problem with this explanation is that M81 stars have also been found, using HST ACS data, in the northern arm (Davidge et al. 2008, Williams et al. 2009), in fact the detection field lies within box 3 shown on Figs. 1. In this case these stars do not seem to give rise to excess far-infrared emission in box 3. So, the position of point 1 on Fig. 2 may be entirely explained by the inherent scatter in the far-infrared HI relation.

\begin{figure}
\centering
\includegraphics[scale=0.55]{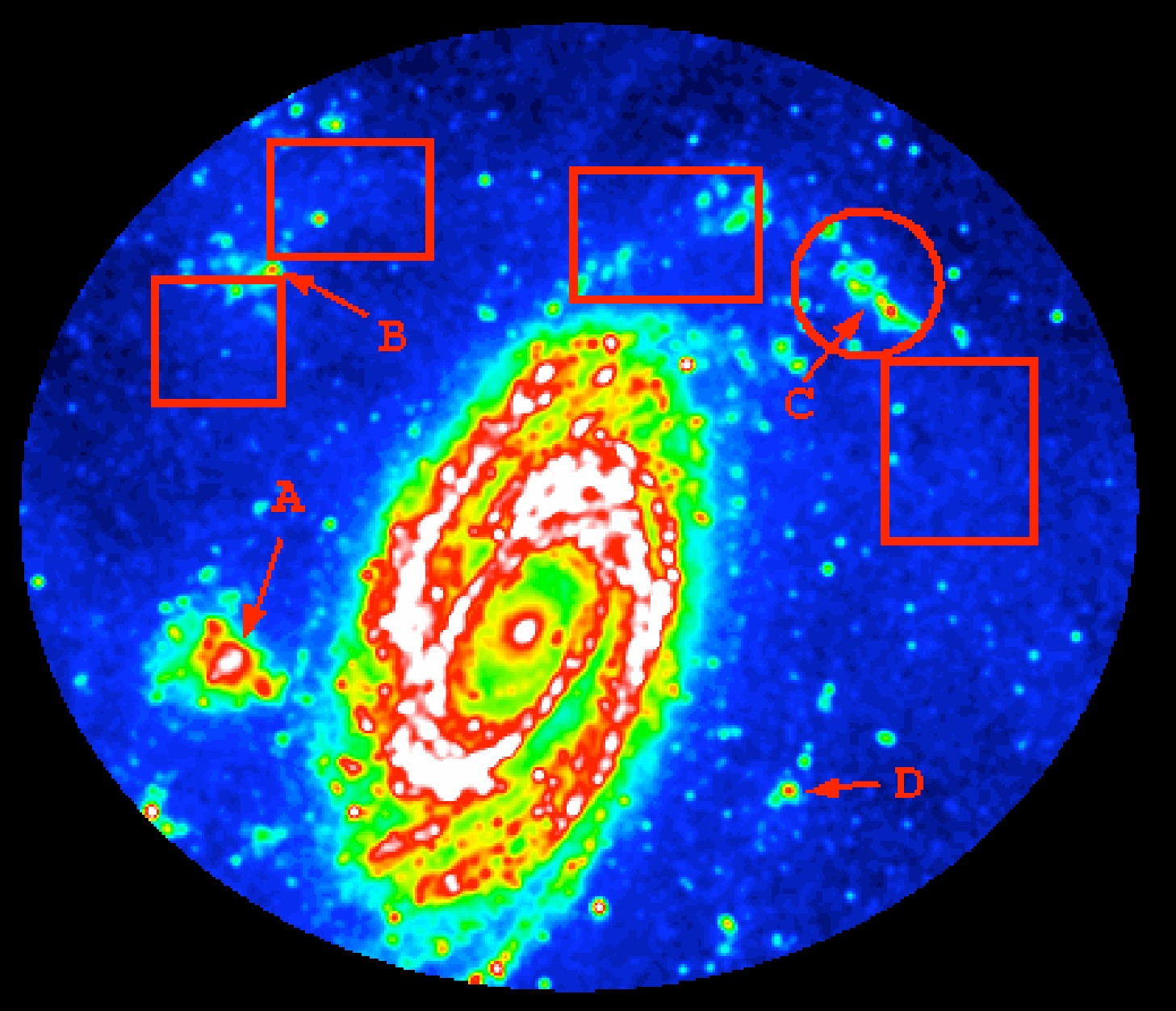}
\includegraphics[scale=0.558]{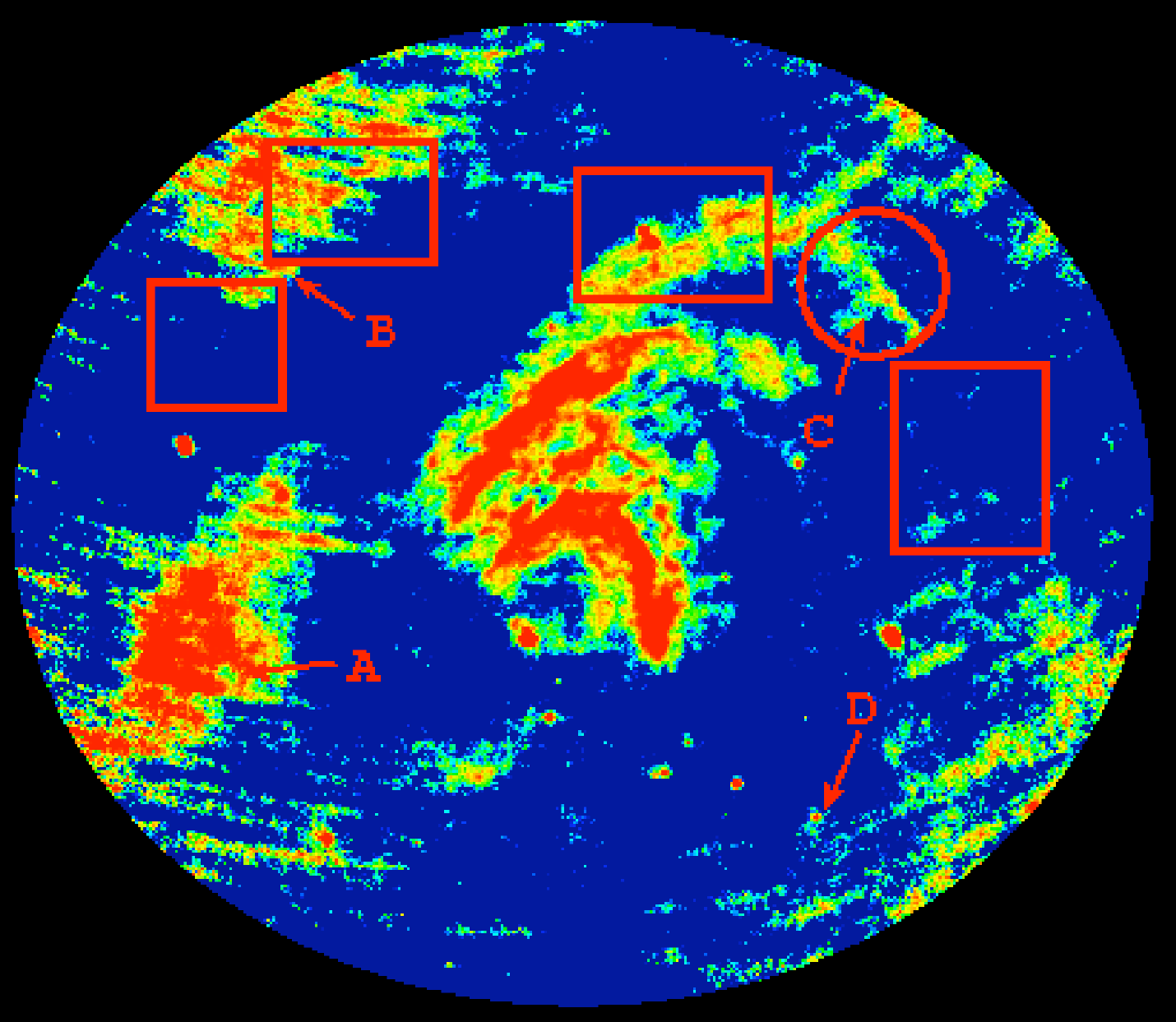}
\caption{Top - the GALEX FUV image of M81, letters indicate A. Holmberg IX, B. A0952+69, C. Northern arm, D. BK 3N. Bottom - THINGS moment zero map summed over 15-180 km s$^{-1}$. The rectangular boxes are the same as those shown on Fig. 1.}
\end{figure}

\section{A closer look at the cirrus}
Using the Herschel and THINGS data we can look at the Galactic cirrus in unprecedented detail (250$\mu$m resolution). Because this is now a study of the inter-stellar medium within our galaxy, which is really beyond the scope of this paper, we will just make some general comments. We concentrate on region 1, but our conclusions are equally applicable to regions 2, 3 and 4. The size of region 1 is approximately 4.5x4.5 sq arc min. We do not know the distance to the cirrus, but if it were at 1 kpc then 4.5$\arcmin$ would correspond to a little over 1 pc and we are investigating the small scale structure of the inter-stellar medium. In Fig. 4a we show a more detailed 250$\mu$m image of region 1 with contours that highlight the brighter regions. The brightest region is labeled A. 

Fig 4b shows the single channel HI emission from Fig 1 with the 250$\mu$m contours overlaid. What is clear is that although there is a good spatial correlation of the far-infrared and HI over scales of order 10$\arcmin$ (Fig 1.) and larger (Boulanger et al. 1996) this relation seems to break down at smaller scales. In the fine detail the far-infrared and HI surface brightness distributions look quite different - note the location of object A and the 'ring' of higher column density in the HI image. Plotting pixel values of the 250$\mu$m emission against the HI leads to a scatter plot considerably more widely dispersed than the Boulanger et al. relation of Fig 2. That the relation is better defined at larger scales is illustrated on Fig. 2 where we mark the position of the mean value of all four boxes - it sits almost exactly on the Boulanger et al. relation, which was derived over scales of 40$\arcmin$. At a level of 3-5 MJy sr$^{-1}$ it is clear that this 'cirrus noise' can be the dominant noise in a far-infrared image. To investigate further the length scales over which the far-infrared and HI spatially correlate we are investigating their cross-power spectrum. This analysis requires fields that are not dominated by a large nearby galaxy but do have high resolution HI. Initial results (Liggins et al., in preparation) indicate that the slope of the cross-power spectrum has a break in it consistent with most of the power of the correlated structures having sizes greater than about 10$\arcmin$.

Fig 4c shows the 250$\mu$m/500$\mu$m colour temperature. The extent of the bright 250$\mu$m emission is again indicated by the contours. Variations in the colour temperature from pixel to pixel across the cirrus, as bounded by the contours, are large compared to the mean temperature fluctuations between individual regions 1-4. The mean flux ratio is 6.5$\pm$1.8 consistent with a temperature of 20$\pm$5K. Similar fluctuations in cirrus temperature, though on half arc minute scales, have been found by Bot et al. (2009). They conclude that these fluctuations must be due to varying dust properties from region to region and not to variations in the inter-stellar radiation field. These small scale variations in dust properties may then also explain the break down of the far-infrared HI correlation. The bright source A is hotter than average having a flux ratio more than twice the mean value, indicating a temperature of order 40K. 

Fig.4d shows a combined SDSS RGB optical image of the same region. This is the diffuse optical emission, mentioned in section 1, which Sun et al. (2005) assign to stars in the M81 group. The far-infrared contours illustrate just how well the optical light traces the dust thermal emission, which in turn seems to be associated with Galactic HI. There are individual optical sources, but because of the far infrared HI association we presume that most of this emission must be back scattered light from stars in the plane of our galaxy.  Just south of source A there is a star, but there also seems to be an optical source almost coincident with it.
The SDSS galaxy J095751.03+691334.1 matches the position of source A to within a pixel. Object A, with a 250$\mu$m flux density of 0.1 Jy, is quite plausibly a background source similar to those found in wide area surveys of the sky (Oliver et al., 2010).

Finally, in Fig 4e we show the GALEX FUV image. There does not appear to be any correlation between the FUV and the far-infrared over these spatial scales. The emission to the north appears to be associated with A0952+69.

\begin{figure*}
\centering
{\Large a)}\includegraphics[scale=0.45]{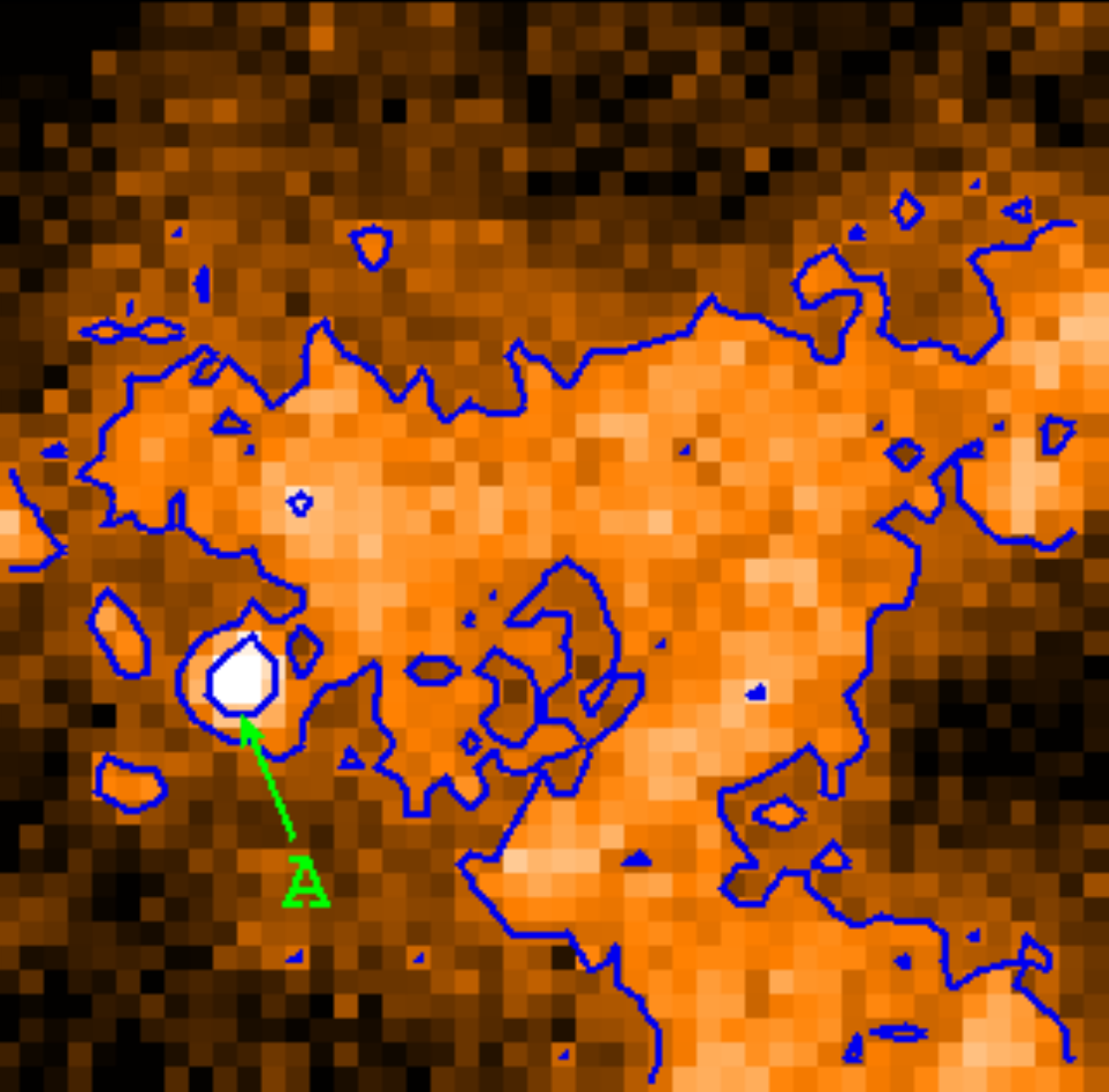}
{\Large b)}\includegraphics[scale=0.45]{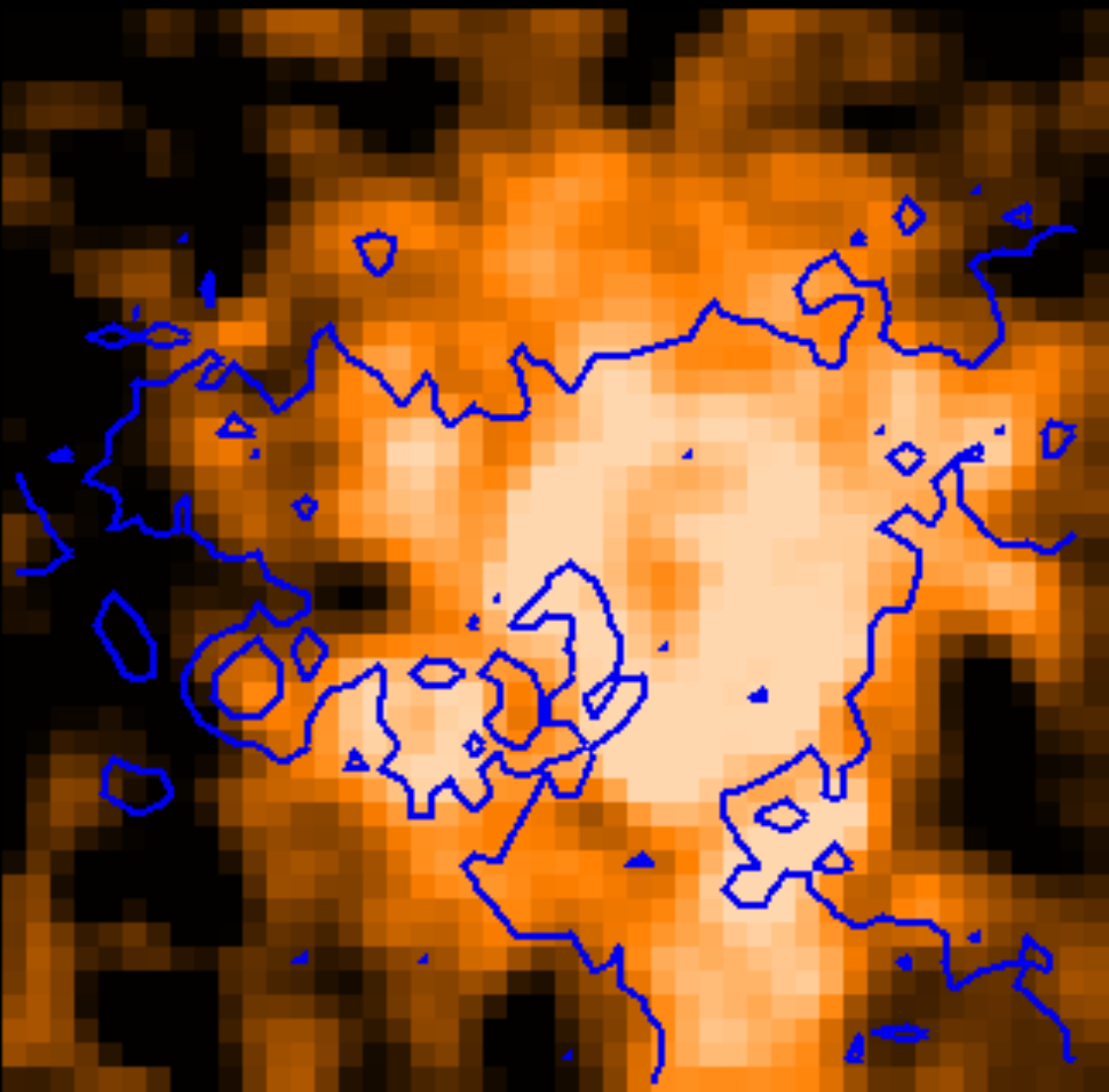}
{\Large c})\includegraphics[scale=0.45]{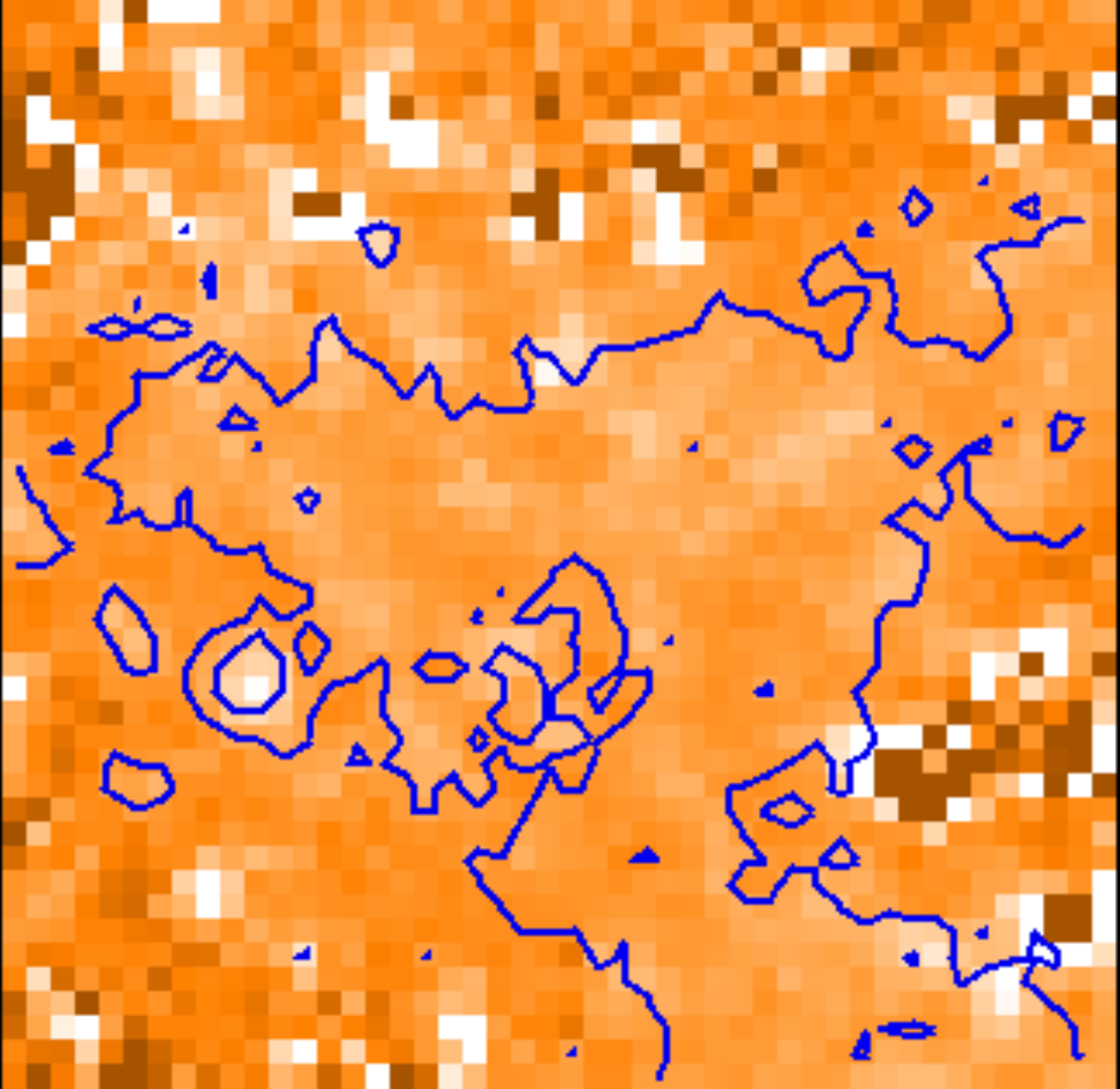}
{\Large d})\includegraphics[scale=0.45]{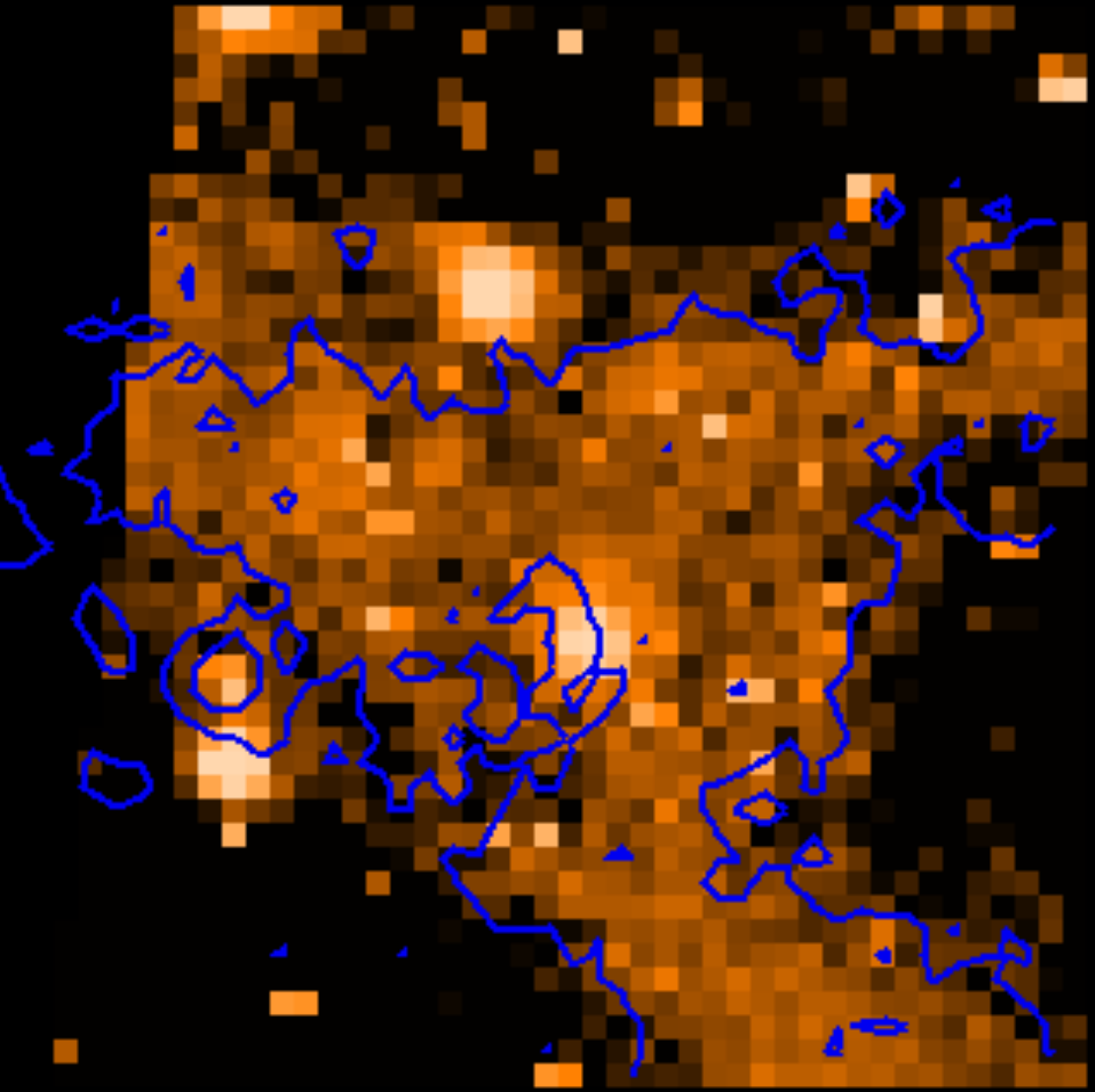}
{\Large e})\includegraphics[scale=0.79]{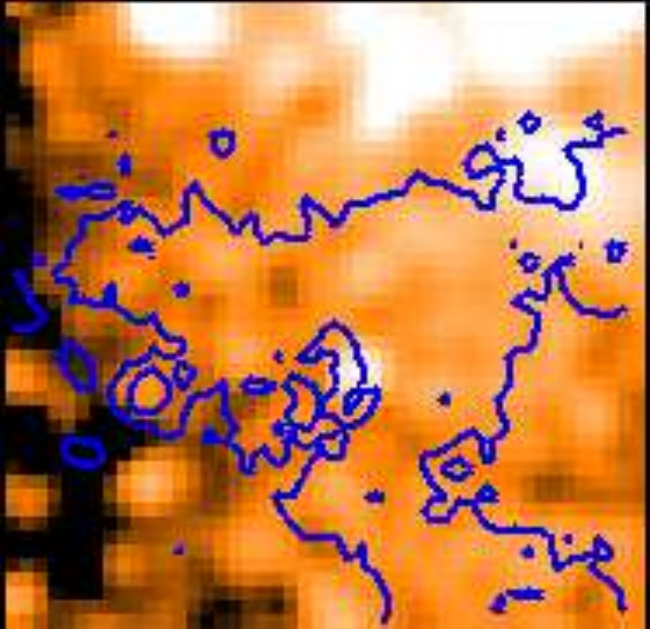}
\caption{a) A zoomed in image of region 1 (250$\mu$m) as defined in Fig 1. It is approximately an area of 4.5x4.5 sq arc min. The contours highlight the regions of brightest emission at a level of 5 and 10 MJy sr$^{-1}$. The brightest source is labeled A. b) The corresponding 21cm emission (channel 75, $v=-1.2$ km s$^{-1}$) with the 250$\mu$m contours. c) The division of the 250$\mu$m image by the 500$\mu$m image with 250$\mu$m contours. d) SDSS optical data with 250$\mu$m contours. e) The FUV image. All images are convolved and re-binned to the 250$\mu$m resolution and beam size.}
\end{figure*}

\section{Conclusions}
Galactic cirrus emission is prominent in all of the Herschel PACS and SPIRE bands making it difficult to unambiguously detect emission from cold diffuse dust in the extra-galactic environment. Some previous measurements of the M81 group's diffuse emission over wavelengths ranging from the optical to the far-infrared have underestimated the significant contribution from Galactic cirrus. We find no evidence for extended dust emission from the M81 group, all of the prominent features can be accounted for by Galactic cirrus. An important result is that not all velocity components of the Galactic cirrus gas have dust associated with them, which is maybe a reflection of the different origins of the gas i.e. infalling primordial gas or expelled enriched disc gas. Also the good relationship between cirrus far-infrared emission and HI seems to break down over spatial scales below about 10$\arcmin$.

Our conclusions provide us with a dilemma at a number of levels. Firstly, the apparent connection of the far-infrared emission with M81, within a galaxy group that has clearly undergone some tidal interaction, makes it very difficult to believe that they are not associated. A particularly nasty astrophysical coincidence! Secondly, given that most of the emission is from Galactic cirrus then what other sources of assumed extra-galactic far-infrared emission may actually also be due to cirrus (Cortese et al. 2010)? Thirdly, there are a number of active star forming regions identified in both the ultra-violet and the HI which have no associated far-infrared emission: far-infrared emission is not a good proxy for star formation in these cases. Fourthly, how can you uniquely distinguish cirrus from extra-galactic emission? Our analysis indicates that you need to select a very narrow velocity range for the HI, but even then there is too much scatter in the Boulanger et al. (1996) relation to be very useful unless you are interested in scales greater than about 10$\arcmin$. This is also true for the variation of far-infrared colours (temperatures) from region to region which can fluctuate considerably over smaller spatial scales.  Fourier filtering (Roy et al. 2010) is a possibility for removing cirrus contamination of deep cosmological surveys that are primarily concerned with point sources (Eales et al. 2010), but for nearby galaxy groups and clusters (Davies et al. 2010) the size of the galaxies (of order $10\arcmin$) compared to the cirrus (Fig. 1) makes this problematic.

\vspace{0.5cm}
\noindent
{\bf ACKNOWLEDGEMENTS} \\
PACS has been developed by a consortium of institutes led by MPE (Germany) and including UVIE 
(Austria); KU Leuven, CSL, IMEC (Belgium); CEA, LAM (France); MPIA (Germany); INAF- 
IFSI/OAA/OAP/OAT, LENS, SISSA (Italy); IAC (Spain). This development has been supported by the 
funding agencies BMVIT (Austria), ESA-PRODEX (Belgium), CEA/CNES (France), DLR (Germany), 
ASI/INAF (Italy), and CICYT/MCYT (Spain).
SPIRE has been developed by a consortium of institutes led by Cardiff University (UK) and including Univ. 
Lethbridge (Canada); NAOC (China); CEA, LAM (France); IFSI, Univ. Padua (Italy); IAC (Spain); 
Stockholm Observatory (Sweden); Imperial College London, RAL, UCL-MSSL, UKATC, Univ. Sussex 
(UK); and Caltech, JPL, NHSC, Univ. Colorado (USA). This development has been supported by national 
funding agencies: CSA (Canada); NAOC (China); CEA, CNES, CNRS (France); ASI (Italy); MCINN 
(Spain); Stockholm Observatory (Sweden); STFC (UK); and NASA (USA).

GALEX (Galaxy Evolution Explorer) is a NASA Small Explorer, launched in April 2003. We gratefully acknowledge NASA's support for construction, operation, and science analysis for the GALEX mission. 

We thank Katie Chynoweth for reprocessing her M81 HI data for us and the THINGS team for providing us with their HI data cube.

\vspace{0.5cm}
\noindent
{\bf REFERENCES} \\
Arp H., 1965, Science, 148, 363 \\
Bendo G. et al., 2010, A\&A, 518, 65 \\
Bot et al., 2009, ApJ, 695, 469 \\
Boulanger F., et al., 1996, A\&A, 312, 256 \\
Chynoweth K., et al., 2008, 135, 1983 \\
Cortese L., et al., 2010, MNRAS, 403, 26 \\
Davidge T., 2008, PASP, 120, 1145 \\
Davies J. et al., 1997, MNRAS, 288, 679 \\
Davies J. et al., 2010, A\&A, 518, 48 \\
Davies J., Alton P., Bianchi S. and Trewhella M., 1998, MNRAS, 300, 1006 \\
del Burgo C., Laureijs, R., Abraham P. and Kiss C., MNRAS, 346, 403 \\
De Mello D., et al., 2008, AJ, 135, 548 \\
Eales et al., 2010, PASP, 122, 499 \\
Greenberg J. M., Ferrini F., Barsella B., Aiello S., 1987, Nature, 327, 214 \\
Griffin M. et al., 2010, A\&A, 518, 3 \\
Hibbard J., et al., 2005, ApJ, 619, L87 \\
Karachentsev I., Karachentseva V., Huchtmeier W. and Makarov D., 2004, AJ, 127, 2031 \\
Li A. and Draine B., 2001, ApJ, 554, 778 \\
Low F. et al., 1984, ApJ, 278, 19  \\
Makarova L., et al., 2002, A\&A, 396, 473 \\
Minchin R. et al., 2007, ApJ, 670, 1056 \\
Mouhcine M. and Ibata R., 2010, MNRAS, in press \\
Oliver S. et al., 2010, A\&A, 518, L210 \\
Poglitsch A. et al., 2010, A\&A, 518, 2 \\
Roy et al., 2010, ApJ, 708, 1611 \\
Sabbi E., et al., 2008, ApJ, 676, L113 \\
Sandage A., 1976, AJ, 81, 954 \\
Smith M. et al., 2010, A\&A, 518, 51 \\
Sollima S. et al., 2010, A\&A, in press \\
Stanimirovic S. et al., 2006, ApJ, 653, 1210 \\
Swaters R. and Balcells M., 2002, AA, 390, 863 \\
Swinyard B. et al., 2010, A\&A, 518, 4  \\
Sun W., et al., 2005, ApJ, 630, L133 \\
Walter F., et al., 2008, AJ, 136, 2563 \\
Williams B., et al., 2009, AJ, 137, 419 \\
Yun M., Ho P., Lo K., 1994, Nature, 372, 530 \\

\end{document}